\documentclass[prl,showpacs,twocolumn,aps]{revtex4}
\usepackage{graphicx}
\usepackage{epsfig}
\usepackage[english]{babel}
\usepackage{amsmath}
\usepackage{amsfonts}
\usepackage{amssymb}
\begin{document}
\title{Logarithmic divergence of the block entanglement entropy for \\
the ferromagnetic Heisenberg model}
\author{Vladislav Popkov}
\affiliation{Institut f\"{u}r Festk\"{o}rperforschung, Forschungszentrum
J\"{u}lich - 52425 J\"{u}lich, Germany.}
\thanks{E-mail: v.popkov@fz.juelich.de}
\author{Mario Salerno}
\thanks{Email: salerno@sa.infn.it}
\affiliation{Dipartimento di Fisica ``E.R. Caianiello" and
Istituto Nazionale di Fisica della Materia (INFM), Universit\'{a}
di Salerno, I-84081 Baronissi (SA), Italy}

\begin{abstract}
Recent studies have shown that logarithmic divergence of
entanglement entropy as function of size of a subsystem is a
signature of criticality in quantum models. We demonstrate that
the ground state entanglement entropy of $ n$ sites for
ferromagnetic Heisenberg spin-$1/2$ chain of the length $L$ in a
sector with fixed magnetization $y$ per site grows as
$\frac{1}{2}\log _{2} \frac{n(L-n)}{L}C(y)$, where $C(y)=2\pi
e(\frac{1}{4}-y^{2})$
\end{abstract}

\maketitle

Recently it has been argued, on the example of the exactly solvable
antiferromagnetic Heisenberg spin $1/2$ chain
\begin{equation}
H_{XXZ}=J\sum_{i{=1}}^{\infty }\left( \sigma _{i}^{x}\sigma
_{i+1}^{x}+\sigma _{i}^{y}\sigma _{i+1}^{y}+\Delta \sigma _{i}^{z}\sigma
_{i+1}^{z}\right) ,  \label{XXZ}
\end{equation}%
that for critical (gapless) quantum system (for the $XXZ$ model
when $\Delta $ belongs to the interval $(-1,1)$) the entanglement
entropy of a block of $n $ spins diverges logarithmically as
$\gamma \log _{2}n$, while for non critical systems ($\Delta$
outside the above mentioned interval), it converges to a constant
finite value \cite{Fazio,Vidal,Latorre_condmat}. This property was
interpreted in the framework of conformal field theory
\cite{Korepin} associated with the corresponding quantum phase
transition and the prefactor $\gamma $ of the logarithm related to
the central charge of the theory $c=3\gamma $ (for the $XXZ$ model
this gives $\gamma =1/3$).

The aim of this Letter is to show that the entanglement entropy of a block
of spins in the ground state of the antiferromagnetic $XXZ$ model (\ref{XXZ}%
), at the point $\Delta=-1$ grows \textit{faster} than for other critical
points $-1<\Delta\leq1$, namely as $\gamma\log_{2}n$ with the logarithmic
prefactor $\frac{1}{2}\leq\gamma\leq1$.

Our approach uses the permutational invariance of the ground state of (\ref%
{XXZ}) at $\Delta=-1$, this allowing to compute the entanglement
entropy exactly for blocks of arbitrary size and system of
arbitrary length. To this regard we remark that by performing the
transformation which overturns each second spin along the chain
(we assume the length of the chain even) the Hamiltonian
(\ref{XXZ}) for $\Delta=-1$ reduces to the isotropic Heisenberg
ferromagnet (\ref{heis}). Since this transformation does not
change the entropy of entanglement, one can compute the block
entropy of the antiferromagnetic Heisenberg chain at $\Delta=-1$
directly from the one of the isotropic ferromagnetic model. It is
worth noting that, in contrast with critical points
$-1<\Delta\leq1$, the point $\Delta=-1$ cannot be studied by means
of conformal field theory since this point is not conformal
invariant \cite{Korepin}, the ground state being infinitely
degenerated at $\Delta=-1$. \cite{Kluemper}

The paper is organized as follows. After introducing the model we
formulate a theorem which gives the analytical expression of the
eigenvalues of the reduced density matrix. Using this theorem we
compute the entanglement entropy of a block of size $n$ in the
finite system of total length $L$ for two specific choices of the
ground state sector. Taking the limit of large subsystem sizes, we
derive analytical expressions for the entanglement entropy
$S_{(n)}$ of a block of spins of size $n$ in the ferromagnetic
ground state, both for $n,L\gg1$ and for $n\gg1,L=\infty$. As a
result, we obtain that in the ground state sector with a fixed
value of $S^{z}$ the block entanglement entropy grows for large
$n$, as $S_{(n)}=\frac{1}{2}\log_{2}\frac{n(L-n)}{L}$, while in
the ground state sector in which all the $S^{z}$ components of the
spin multiplet are equally weighted, $S_{(n)}=\log_{2}(n+1)$ for
arbitrary $n$ and $L$.

We consider the ferromagnetic Heisenberg model with nearest
neighbor interaction,
\begin{equation}
H_{XXX}=-J\sum_{i{=1}}^{L}\left( \overset{\rightarrow }{\sigma _{i}}\overset{%
\rightarrow }{\sigma _{i+1}}-3I\right)   \label{heis}
\end{equation}%
where $\sigma $ are Pauli matrices, $J>0$ denotes the exchange constant and $%
L$ the number of spins (we assume periodic boundary conditions $L+1\equiv 1$%
). As is well known, the ground state of (\ref{heis}) belongs to a multiplet
of total spin $S=\frac{L}{2}$ and is degenerate with respect to $S^{z}=-%
\frac{L}{2},-\frac{L}{2}+1,...\frac{L}{2}$ . In the sector with a fixed
number $N$ of spins down, i.e. with a fixed $S^{z}=N-\frac{L}{2}$, the
ground state is obtained by the action of the rising operator $%
S^{+}=\sum_{i}\sigma _{i}^{+}$ on the vacuum state with all spins down
\begin{equation}
|\Psi _{L}^{N}\rangle \sim (S^{+})^{N}|\downarrow \downarrow ...\downarrow
\rangle .  \label{spin_GS}
\end{equation}%
All eigenfunctions (\ref{spin_GS}) correspond to the same ground state
energy $E=0$ of the $XXX$ model (\ref{heis}). The structure of the state (%
\ref{spin_GS}) is given by
\begin{equation}
|\Psi (L,N)\rangle =\frac{1}{\sqrt{C_{N}^{L}}}\sum_{P}|\underbrace{\uparrow
\uparrow ...\uparrow }_{N}\underbrace{\downarrow \downarrow ...\downarrow
\rangle }_{L-N}  \label{GS}
\end{equation}%
where the sum is taken over all possible distributions of $N$ spins on $L$
sites and the binomial coefficient $C_{N}^{L}=\frac{L!}{N!(L-N)!}$ takes
care of the normalization. Note that (\ref{GS}) is also a ground state for
the model of interacting bosons \cite{Mario}, while for the partially
asymmetric exclusion process ASEP \cite{Schu00} with $N$ particles hopping
with hard-core exclusion on a closed chain of the length $L$, (\ref{GS})
represents a steady-state vector. We will be interested in the ground state
entanglement (von Neumann) entropy $S_{(n)}$ of a block of $n$ (not
necessarily contiguous) spins
\begin{equation}
S_{(n)}=-tr(\rho _{n}\log _{2}\rho _{n})=-\sum \lambda _{k}\log _{2}\lambda
_{k},  \label{entropy}
\end{equation}%
where $\rho_{n}$ is the reduced density matrix of the block,
obtained from the density matrix $\rho $ of the whole system by
tracing out external degrees of freedom $\rho_{(n)}=tr_{(L-n)}\rho
$ (notice that due to the permutational symmetry of the ground
state $S_{(n)}$ does not depend on the particular choice of the
block but only on its size $n$). In Eq. (\ref{entropy}) $\lambda
_{k}$ are the eigenvalues of the reduced density matrix which are
all real, nonnegative, and sum up to one: $\sum \lambda _{k}=1$.

The density matrix $\rho$ for a degenerate ground state is given by
\begin{equation}
\rho=\sum_{N=0}^{L}\alpha_{N}|\Psi(L,N)\rangle\langle\Psi(L,N)|,\text{ \ \ \
\ }\;\sum\alpha_{N}=1\text{,}
\label{BIG_rho_general}
\end{equation}
where $\alpha_{0},\alpha_{1},...\alpha_{L},$ is a set of nonnegative
coefficients. Denoting the reduced density matrix in a fixed sector with $N$
spins up by $\rho_{n}(N)$,
\begin{equation}
\rho_{n}(N)=tr_{(L-n)}|\Psi_{(}L,N)\rangle\langle\Psi(L,N)|,
\label{reduced_density_matrix_N}
\end{equation}
where $|\Psi(L,N)\rangle$ is given by (\ref{GS})), one can write the general
reduced density matrix as
\begin{equation}
\rho_{n}=\sum_{N=0}^{L}\alpha_{N}\text{ }\rho_{n}(N).
\label{reduced_density_matrix_general}
\end{equation}
In the following we consider two choices for the coefficients $\{\alpha_{i}\}
$:
\begin{align}
& \text{\textbf{(a)}}\;\;\;\;\;\;\;\alpha_{i}=\delta_{iN},
\label{maximally_anisotropic_set} \\
&
\text{\textbf{(b)}}\;\;\;\;\;\;\;\alpha_{0}=\alpha_{1}=...=\alpha
_{L}= \frac{1}{L+1}   \label{isotropic_set}
\end{align}
(the analysis for arbitrary $\{\alpha_{i}\}$ proceeds in similar
manner). The choice $(a)$ corresponds to the case when a small
anisotropy single out a sector with $N$ spins up resulting in a
pure state of a global system, see
(\ref{GS},\ref{BIG_rho_general}). The choice $(b)$ corresponds to
an equilibrated density matrix (i.e. with all components of the
ground state multiplet equally weighted) which preserves the
$SU(2)$ invariance of the Hamiltonian (\ref{heis}) (this case is
equivalent to infinite temperature). Using the general property of
the entropy of composite systems: $S_{(n)}=S_{(L-n)}$, and its
invariance with respect to the inversion of all spins, we can
restrict the analysis, without loosing generality, to the case:
$n\leq \frac{L}{2},\;N\leq \frac{L}{2}$. The computation of the
block entanglement entropy is drastically simplified by the
following

\noindent\textbf{Theorem}: \textit{The eigenvalues of the reduced density
matrix }$\rho_{n}(N)$\textit{\ of a block of }$n$\textit{\ spins in the
sector with }$N$\textit{\ spins up in the ground state of the ferromagnetic
Heisenberg model (\ref{heis}) are given by}
\begin{equation}
\lambda_{k}(L,n,N)=\frac{C_{k}^{n}C_{N-k}^{L-n}}{C_{N}^{L}}%
,\;\;\;k=0,1,...\min(n,N).   \label{eigenvalues_finite}
\end{equation}
The proof of the theorem follows from the decomposition of $\rho_{n}(N)$
with respect to the symmetric orthogonal subspaces of the system of $n$
spins, classified by the integer $k=0,1,...\min(n,N)$ giving the number of
spins up in the block
\begin{equation}
\rho_{n}(N)=\sum_{k=0}^{\min\{n,N\}}c_{k}|\psi(n,k)\rangle\langle\psi(n,k)|.
\label{decomposition}
\end{equation}
Here $|\psi(n,k)\rangle$ denotes the symmetric state with $k$ spins up among
$n$ spins
\begin{equation}
|\psi(n,k)\rangle=\sum_{P}|\underbrace{\uparrow\uparrow...\uparrow}_{k}%
\underbrace{\downarrow\downarrow...\downarrow}_{n-k}\rangle
\end{equation}
and $c_{k}$ is the corresponding probability $c_{k}=\frac{C_{N-k}^{L-n}}{%
C_{N}^{L}}$ (notice that $C_{N-k}^{L-n}$ is the number of states
with $k$ spin up in the block of $n$ spins and ${C_{N}^{L}}$ is
the total number of states). Expression (\ref{decomposition}) can
be rewritten as
\begin{equation}
\rho_{n}(N)=\sum_{k=0}^{\min\{n,N\}}\lambda_{k}\rho_{n}(k)
\label{decomposition2}
\end{equation}
where $\rho_{n}(k)$ is the density matrix of the state
$|\psi(n,k)\rangle$ and the coefficients
$\lambda_{k}=\frac{C_{k}^{n}C_{N-k}^{L-n}}{C_{N}^{L}}$ sum up to
one, $\sum\lambda_{k}=1$. From this it follows that $\rho_{n}(N)$
is the density matrix associated with the ensemble of orthogonal
pure states $\{\lambda_{k},\rho_{n}(k)\}$ and therefore it has a
block diagonal form, each block having only one nonzero eigenvalue
$\lambda_{k}$ which coincides with the expression
(\ref{eigenvalues_finite}). This concludes the proof of the
Theorem.

We remark that the specific case $N=n=\frac{L}{2}$ was also
considered in Ref. \cite{peshel}. Having found the eigenvalues of
$\rho_{n}(N)$ one can easily compute the entanglement entropy
$S_{(n)}$ for arbitrary $L,n$ and $N$.

\noindent

\paragraph{\textbf{Case (a)}}

\begin{figure}[tbp]
\centerline{\includegraphics[width=8.15cm,height=8.cm,clip]{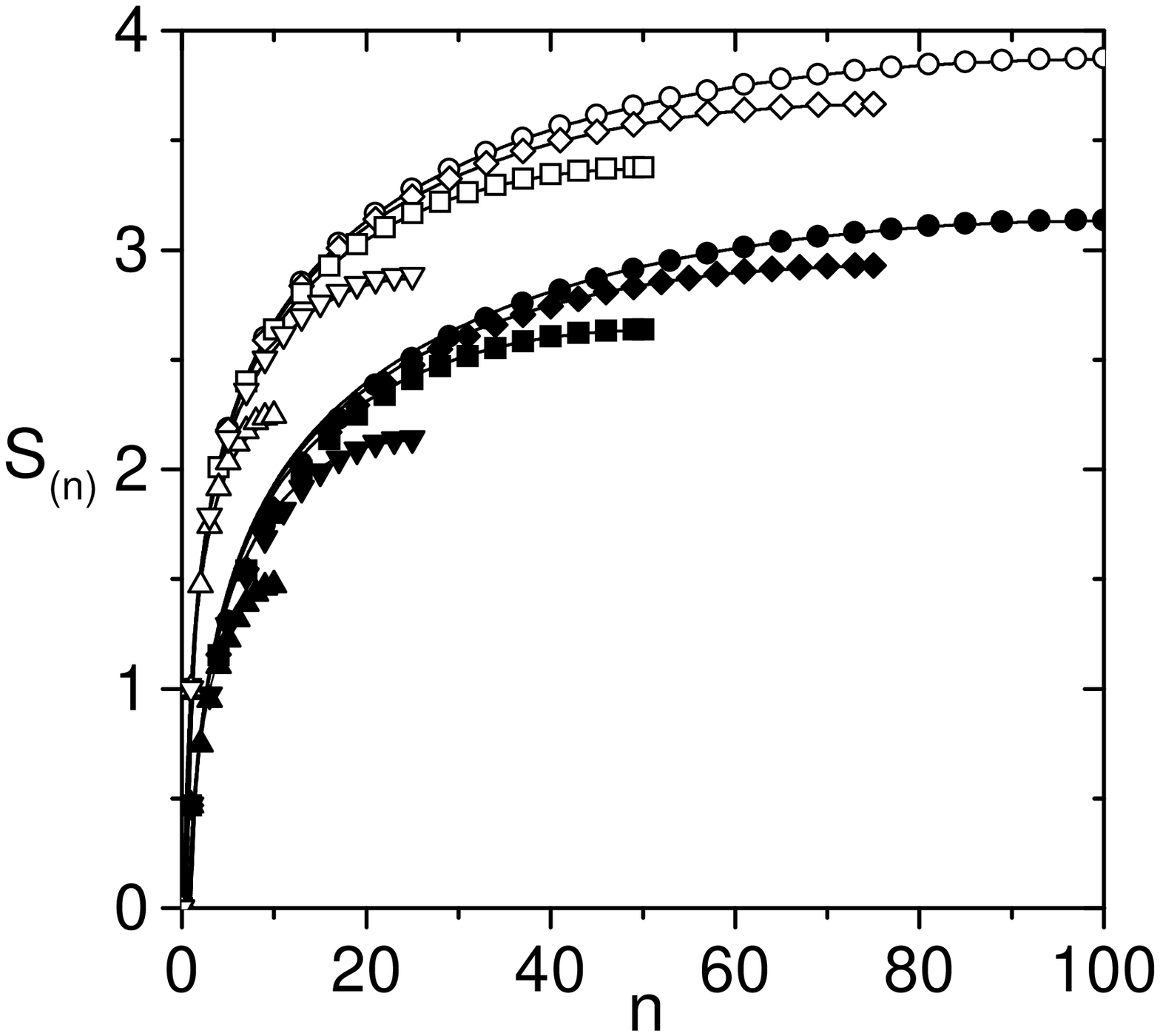}}
\caption{Entanglement entropy as obtained from the exact
expressions Eqs.
(\protect\ref{entropy},\protect\ref{eigenvalues_finite}), as a
function of the block size $n$ and for $L=20$ (up triangles), $50$
(down triangles), $100 $ (squares), $150$ (diamonds), $200$
(circles). Filled (empty) symbols correspond to  $p=1/10$
($p=1/2$). Continuous curves represent the
analytical expression in Eq. (\protect\ref{entropy_L}). For $n>L/2$, $%
S_{(n)}=S_{(L-n)}$(not shown).}
\label{fig_entropy_L}
\end{figure}

To obtain an analytical expression for $S_{(n)}$, from the exact
expression (\ref{entropy}, \ref{eigenvalues_finite}), we observe
that for blocks of large size, $n\gg 1$, the dominant contribution
to the sum (\ref{entropy}) comes from the eigenvalues $\lambda
_{k}$ with large $k$. In this case one can approximate the
binomial coefficients in (\ref{eigenvalues_finite}) by the normal
distribution, see e.g. \cite{Allen}:
\begin{equation}
C_{n}^{m}p^{m}q^{n-m}\approx \frac{1}{\sqrt{2\pi npq}}\exp \left( -\frac{%
(m-np)^{2}}{2npq}\right) ,\text{ \ \ }npq\gg 1,  \label{binom_approx}
\end{equation}%
where $0<p<1,q=1-p$ . Using this approximation, and defining $p=N/L$, the
eigenvalues (\ref{eigenvalues_finite}) can be written as
\begin{align*}
\lambda _{k}(L,n,N)& =\frac{%
C_{k}^{n}p^{k}q^{n-k}C_{N-k}^{L-n}p^{N-k}q^{L-n-N+k}}{C_{N}^{L}p^{N}q^{L-N}}
\\
& \approx \frac{1}{n}\frac{1}{\sqrt{2\pi \alpha }}\exp \left( -\frac{(\frac{k%
}{n}-p)^{2}}{2\alpha }\right) ,
\end{align*}%
where $\alpha =\frac{pq(L-n)}{nL}$. Substituting this expression into (\ref%
{entropy}) and replacing the sum with an integral, we obtain
\begin{align*}
& S_{(n)}(p)\approx \int_{0}^{1}R\left( \log _{2}{\frac{R}{n}}\right) dx, \\
& R=\frac{1}{\sqrt{2\pi \alpha }}\exp \left( -\frac{(x-p)^{2}}{2\alpha }%
\right) .
\end{align*}%
For large $n$ the limits of the integral can be extended to
include the whole real axis, after which the result of the
integration gives
\begin{equation}
S_{(n)}(p)\approx \frac{1}{2}\log _{2}(2\pi epq)+\frac{1}{2}\log
_{2}{\frac{n(L-n)}{L}}. \label{entropy_L}
\end{equation}
Notice that this approximate result is valid for $npq\gg 1$ and in
the limit $npq\rightarrow \infty $ it becomes exact. From the
analytical expression (\ref{entropy_L}) the following properties
can be easily derived: \textit{i)} $S_{(n)}(p)=S_{(n)}(1-p)$,
\textit{ii)} $\ S_{(n)}(p)=S_{(L-n)}(p)$, \textit{ iii)} $\partial
S_{(n)}(p)/\partial n=0$ only at $n=\frac{L}{2}$, \textit{vi)}
$\partial S_{(n)}(p)/\partial p=0{\ }$ only at $p=\frac{1}{2}$,
\textit{v)} $S_{(n)}(p)$ is a monotonically increasing function of
the total length $L$. In Fig.  (\ref{fig_entropy_L}) we compare
the exact entropy of finite systems, as computed from exact
expressions Eqs. (\ref{entropy}, \ref{eigenvalues_finite}), with
the analytical expression (\ref{entropy_L}), from which we see
that there is an excellent agreement also for small values
of $npq$. In the thermodynamic limit $L\rightarrow \infty ,\frac{N}{L}%
\rightarrow p$ the eigenvalues (\ref{eigenvalues_finite}) reduce to
\begin{equation}
\lambda _{k}=C_{n}^{0}p^{n},C_{n}^{1}p^{n-1}q,...C_{n}^{n}q^{n},
\label{eigenvalues_thermodynamic}
\end{equation}
and the corresponding entanglement entropy is obtained from
(\ref{entropy_L}) as
\begin{equation}
S_{(n)}(p)\approx \frac{1}{2}\log _{2}(2\pi epq)+\frac{1}{2}\log _{2}n.
\label{entropy_n(p)_limit}
\end{equation}

\begin{figure}[tbp]
\begin{center}
\centerline{%
\includegraphics[width=8.15cm,height=8.3cm,clip]{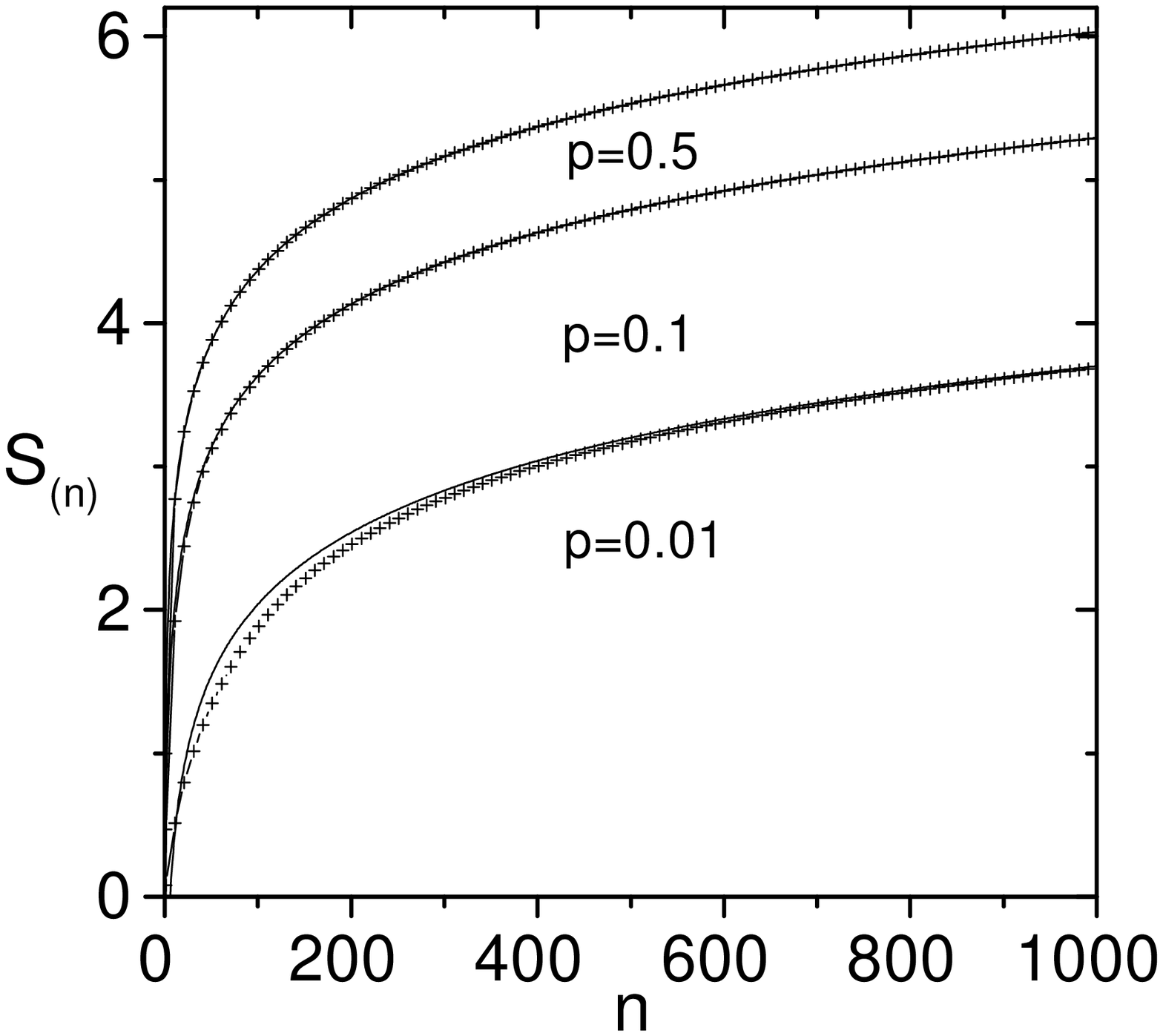}}
\end{center}
\caption{Entanglement entropy as function of a number of sites
involved, for different values of
$p=\frac{1}{100},\frac{1}{10},\frac{1}{2}$. Comparison of exact
formula (points) with the limiting expression
(\protect\ref{entropy_n(p)_limit}) (continuous curves). }
\label{fig_entropy_new}
\end{figure}

In Fig. \ref{fig_entropy_new} we plot the exact entanglement
entropy of a block of size $1\leq n\leq 1000$ in an infinite chain
(\ref{entropy}), (\ref{eigenvalues_thermodynamic}), versus the
limiting expression (\ref{entropy_n(p)_limit}) for different
filling $p$.
We see that the analytic formula (\ref{entropy_n(p)_limit}) gives
a good approximation even for small finite number of sites $n$ in
the block. For very small $p$ the convergence is slower (see the
lowest graph in Fig. \ref{fig_entropy_new}) because the validity
of formula (\ref{binom_approx}) crucially depends on the value of
$npq$.

Thus, for case $(a)$ we conclude that the block entanglement entropy of the
ferromagnetic ground state grows logarithmically with $n$, as for critical
quantum systems, but with a different prefactor, i.e., as $\frac{1}{2}%
\log_{2}n$ rather than $\frac{1}{3}\log_{2}n$ predicted in \cite%
{Latorre_condmat}.

\noindent

\paragraph{\textbf{Case (b)}}

In this case the eigenvalues of the reduced density matrix are given by
\begin{equation}
\lambda_{k}=\frac{C_{n}^{k}}{L+1}\sum_{N=n-k}^{L-k}\frac{C_{L-n}^{N-n+k}}{%
C_{L}^{N}}=\frac{1}{n+1},\;k=0,1,....n   \label{GS_eigenvalues_isotropic}
\end{equation}
and are independent on $k$ and on the size of the system $L$. The
entanglement entropy is obtained as
\begin{equation}
S_{(n)}=\log_{2}(n+1),\;n=1,2,....L.   \label{entropy_isotropic}
\end{equation}
Equations
(\ref{eigenvalues_finite},\ref{entropy_L},\ref{entropy_n(p)_limit}
) and (\ref{GS_eigenvalues_isotropic},\ref{entropy_isotropic}),
corresponding to the cases (a) and (b) considered above are the
main results of the paper.

It is worth to note that, due to the permutational invariance of
the ground state, for any choice of the density matrix
(\ref{reduced_density_matrix_general}) the reduced density matrix
for a block of size $n$ has exactly $n+1$ nonzero eigenvalues (see
the theorem) in the ground state. This implies the upper bound for
the entropy $S_{\max }(n)=\log _{2}(n+1)$, which is achieved in
the case of a thermally equilibrated density matrix (case (b)).
The lower bound of logarithmic growth $S_{(n)}\sim \frac{1}{2}\log
_{2}n$ \ is achieved for the \textquotedblleft
anysotropic\textquotedblright\ choice corresponding to a pure
state (\ref{BIG_rho_general}) of the whole system (case (a)). For
generic choice of the coefficients $\{\alpha _{N}\}$ in
(\ref{BIG_rho_general},\ref{reduced_density_matrix_general}),
$S_{(n)}$ will grow as $\gamma \log _{2}n$ with $\frac{1}{2}\leq
\gamma $ $\leq 1$.

We also note that (\ref{entropy_isotropic}) is a monotonically
increasing function of $n$, attaining maximum for the whole system
$n=L$, while in the case of pure state the maximum is achieved for
a block of half-system size $n=\frac{L}{2} $. This feature is
related to the fact that the ground state of a ferromagnet is
highly degenerate and the total system for the choice
(\ref{isotropic_set}) is in the maximally mixed state.

Another remark concerns the origin of the logarithmic prefactor
$\gamma = \frac{1}{2}$ in formula (\ref{entropy_n(p)_limit}).
Apparently $\gamma $ is not related to any central charge in since
$\Delta =-1$ is not a conformal point. We find that in our case
the prefactor $\gamma $ is related to the spin $s$ per site, i.e.,
one can show that for a ferromagnetic spin $s$ chain (i.e. with
on-site spin $s$), the block entanglement entropy in the ground
state sector grows like $S_{(n)}\simeq const+s\log _{2}n$ (details
will be presented elsewhere). We finally remark that it is of
interest to generalize
Eqs.(\ref{entropy_L},\ref{entropy_isotropic}) to the case of
nonzero temperature, where excited states have to be taken into
account. Work in this direction is in progress.

\acknowledgments The authors wish to thank V. Korepin, G.
Sch\"{u}tz, I. Peschel, and A. Kl\"{u}mper for interesting
discussions. VP acknowledges the INFM, Unit\'{a} di Salerno, for
providing a three months grant during which this work was done,
and the Department of Physics of the University of Salerno for the
hospitality. MS acknowledges partial financial support from
J\"{u}lich-Forschungszentrum and from a MURST-PRIN-2003
Initiative. 

\end{document}